\documentclass[copyright,creativecommons]{eptcs}

\providecommand{\event}{M. Bujorianu and M. Fisher (Eds.):\\
\hbox{\quad}Workshop on Formal Methods for Aerospace (FMA)}
\providecommand{\volume}{20}
\providecommand{\anno}{2010}
\providecommand{\firstpage}{1}
\providecommand{\eid}{1}

\usepackage{amsthm,mathptmx,amsfonts,amssymb,amsmath}
\usepackage{graphics}

\title{Bisimulation Relations Between Automata, \\
Stochastic Differential Equations and Petri Nets}

\author{Mariken H.C. Everdij \qquad\qquad Henk A.P. Blom
\institute{National Aerospace Laboratory NLR\\ Amsterdam, Netherlands}
\email{everdij@nlr.nl \qquad\qquad blom@nlr.nl}
}
\def\titlerunning{Bisimulation Relations Between Automata, Stochastic Differential Equations and Petri Nets}
\def\authorrunning{M.H.C. Everdij \& H.A.P. Blom}

\begin{document}

\maketitle

\begin{abstract}
Two formal stochastic models are said to be bisimilar if their
solutions as a stochastic process are probabilistically
equivalent. Bisimilarity between two stochastic model formalisms
means that the strengths of one stochastic model formalism can be
used by the other stochastic model formalism. The aim of this
paper is to explain bisimilarity relations between stochastic
hybrid automata, stochastic differential equations on hybrid space
and stochastic hybrid Petri nets. These bisimilarity relations
make it possible to combine the formal verification power of
automata with the analysis power of stochastic differential
equations and the compositional specification power of Petri nets.
The relations and their combined strengths are illustrated for an
air traffic example.
\end{abstract}

\section{Introduction}\label{sec:introduction}
Two formal stochastic models are said to be
\emph{bisimilar} if their solutions as a stochastic process (i.e.\
their \emph{executions}) are probabilistically equivalent
\cite{BujorianuLygerosBujorianu2005,VanderSchaft2004}.
Bisimilarity relations between formal stochastic models are very
useful to study since they allow one stochastic model to take advantage
of the strengths of the other stochastic model. The aim of this paper is
to show bisimulation relations between three different stochastic modelling
formalisms: stochastic hybrid automata, stochastic differential
equations on hybrid space, and stochastic hybrid Petri nets. These bisimulation relations make it
possible to combine the formal verification power of automata with
the analysis power of stochastic differential equations and
the compositional specification power of Petri nets.

For the stochastic automata formalism, we take the \emph{general stochastic
hybrid system} (GSHS) theoretical setting developed by
\cite{BujorianuLygeros2006}. A GSHS is a hybrid automaton defined
on a hybrid state space. This hybrid state space consists of a
countable set of discrete modes, and per discrete mode a Euclidean
subset. Per discrete mode, a stochastic differential equation (SDE)
is defined. Two additional GSHS elements are a jump rate function and
a GSHS transition measure. The execution of these elements
provides a stochastic process that follows the solution of the
SDE connected to the initial discrete
mode. After a time period, defined by the jump rate function, the
process state may spontaneously jump to another mode, defined by
the GSHS transition measure. A jump may also be forced if the
process state hits the boundary of the Euclidean subset. The GSHS
execution is referred to as \emph{general stochastic hybrid
process} (GSHP). One of the main strengths of the automata
formalism is the availability of formal verification tools.

For the hybrid stochastic differential equations formalism we take the
\emph{hybrid stochastic differential equations} (HSDE) theoretical
setting developed in a series of complementary studies
\cite{Blom2003,BlomBakkerEverdijPark2003b,Krystul2006,KrystulBlomBagchi2007}.
A HSDE consists of a sequence of SDEs
on a hybrid state space, driven by a Poisson random measure. When
the Poisson random measure generates a multivariate point, a
spontaneous jump occurs. A jump may also be forced if the process
state hits the boundary of a Euclidean subset. The HSDE solution
process is referred to as \emph{general stochastic hybrid process}
(GSHP). In \cite{EverdijBlom2009} it is shown that whereas the
GSHS formalism is at some points more general than HSDE (for GSHS
the dimension of the Euclidean subset may depend on the discrete
mode; for HSDE this dimension is fixed), HSDE has the advantage of
an established semi-martingale property and includes the coverage
of jump-linear systems.

For the stochastic hybrid Petri nets formalism, we take \emph{stochastically and
dynamically coloured Petri nets} (SDCPN) developed in a series of
studies by
\cite{EverdijBlom2003,EverdijBlom2005,EverdijBlom2006}. A Petri
net has \emph{places} (circles), which model possible discrete
states or conditions, and which may contain one or more
\emph{tokens} (dots), modelling which of these states are current.
The places are connected by \emph{transitions} (squares), which
model state switches by removing input tokens and producing output
tokens along \emph{arcs} (arrows). In SDCPN, the tokens have
Euclidean-valued colours that follow SDEs. Some of the transitions
remove and produce tokens
spontaneously, other transitions are forced and occur
when the colours of their input tokens reach the boundary of a
Euclidean subset. The collection of token colours in all places
forms a \emph{general stochastic hybrid process} (GSHP). The
specific strength of SDCPN is their compositional specification
power, which makes available a hierarchical modelling approach
that separates local modelling issues from global modelling
issues. This is illustrated for a large distributed example in air
traffic management \cite{EverdijKlompstraBlomKleinObbink2006},
which covers many distributed agents each of which interacts in a
dynamic way with the others. Other typical Petri net features are
concurrency and synchronisation mechanism, hierarchical and
modular construction, and natural expression of causal
dependencies, in combination with graphical and equational
representation.

\begin{figure}[h!bt]
\begin{center}
\includegraphics{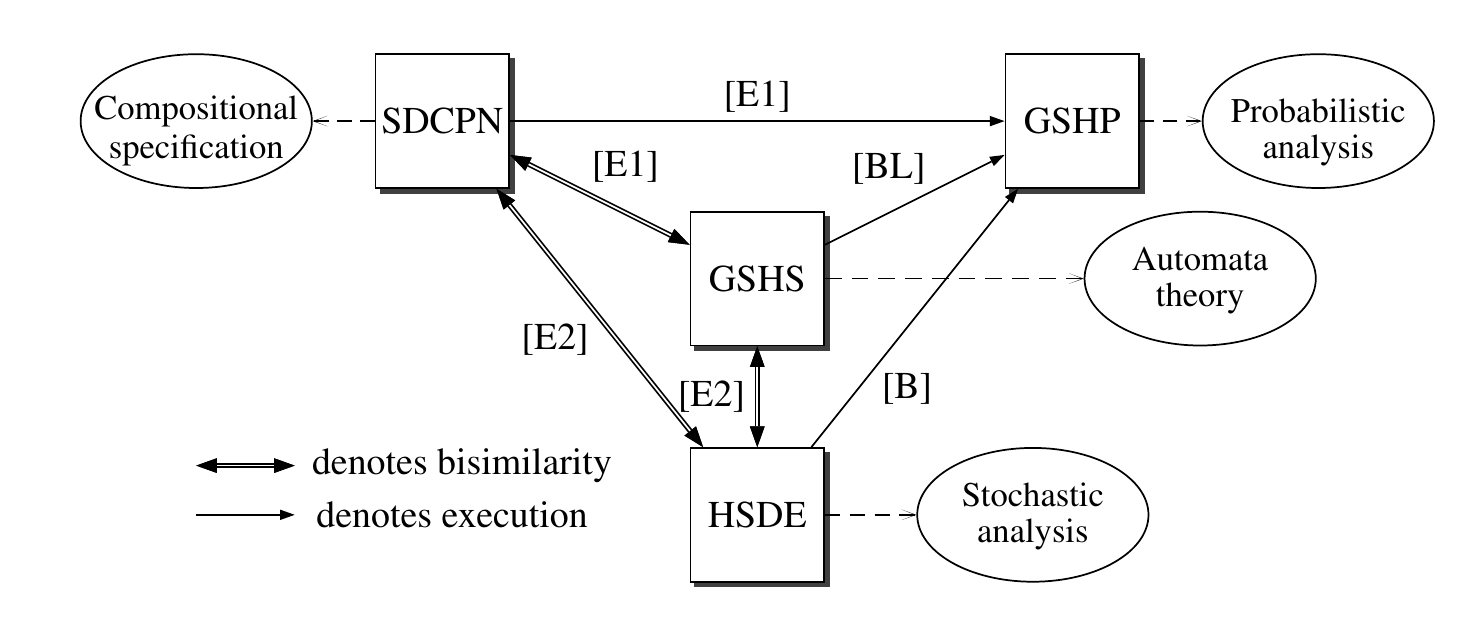}
\end{center}
\caption{Relationship between SDCPN, GSHS, GSHP and HSDE, and
their key properties and advantages.
The [B] arrow is established in \protect\cite{Blom2003}.
The [BL] arrow is established in \protect\cite{BujorianuLygeros2006}.
The [E1] arrows are established in \protect\cite{EverdijBlom2006}.
The [E2] arrows are established in \protect\cite{EverdijBlom2009}.}
\label{fig:GSHS:GSHP:SDCPN}
\end{figure}

The aim of this paper is to illustrate the relations between
SDCPN, GSHP, HSDE and GSHS which show that SDCPN, GSHS and HSDE
are bisimilar. This means that if we take the elements of any one
of these formalisms, we can construct the elements of another
formalism in such a way that their associated GSHPs are
probabilistically equivalent. Fig.\ \ref{fig:GSHS:GSHP:SDCPN}
shows the relations between the formalisms, and the key tools
available for each of them.

With these relations, the properties and advantages of the various
approaches come within reach of each other. The compositional
specification power of SDCPN makes it relatively easy to develop a
model for a complex system with multiple interactions.
Subsequently, in the analysis stage three alternative approaches
can be taken. The first is direct execution of SDCPN and
evaluation through e.g.\ Monte Carlo simulation. The second is
mapping the SDCPN into a GSHS and evaluating its execution, with
the advantages of connection to formal methods in automata theory
and to optimal control theory \cite{BujorianuLygeros2004cdc}. The
third is mapping the SDCPN into HSDE and evaluating its solution,
with the advantages of stochastic analysis for semi-martingales
\cite{Elliott1982,ElliottAggounMoore1995}. With the GSHP
resulting from any of these three means, properties become
available such as convergence of discretisation, existence of
limits, existence of event probabilities, strong Markov
properties, and reachability analysis
\cite{BujorianuLygeros2006,Davis1993,EthierKurtz1986}.

The organisation of this paper is as follows. Section
\ref{sec:sdcpn} defines SDCPN and the related SDCPN process.
Section \ref{sec:airtrafficSDCPN} presents an example SDCPN model
for a simple but illustrative air traffic situation. Section
\ref{sec:gshsautomaton} defines GSHS and illustrates how the
example SDCPN can be mapped to a bisimilar GSHS. Section
\ref{sec:hsde} defines HSDE and illustrates how the example SDCPN
can be mapped to a bisimilar HSDE. Section \ref{sec:conclusions}
gives conclusions.

\section{SDCPN}\label{sec:sdcpn}
This section outlines \emph{stochastically and
dynamically coloured Petri net} (SDCPN). For a more formal
definition, we refer to \cite{EverdijBlom2009}.

\paragraph{Definition 2.1} (Stochastically and dynamically coloured Petri net.)
\emph{An SDCPN is a collection of elements $({\cal P}$, ${\cal T}$,
${\cal A}$, ${\cal N}$, ${\cal S}$, ${\cal C}$, ${\cal I}$, ${\cal
V}$, ${\cal W}$, ${\cal G}$, ${\cal D}$, ${\cal F})$, together
with an SDCPN execution prescription which makes use of a sequence
$\{U_i ; i = 0, 1, \ldots \}$ of independent uniform $U[0,1]$
random variables, of independent sequences of mutually independent
standard Brownian motions $\{B_{t}^{i, P} ; i = 1, 2, \ldots \}$
of appropriate dimensions, one sequence for each place $P$, and of
five rules R0--R4 that solve enabling conflicts.}

\subsection{SDCPN elements}
\begin{figure}[h!bt]
\begin{center}
\includegraphics{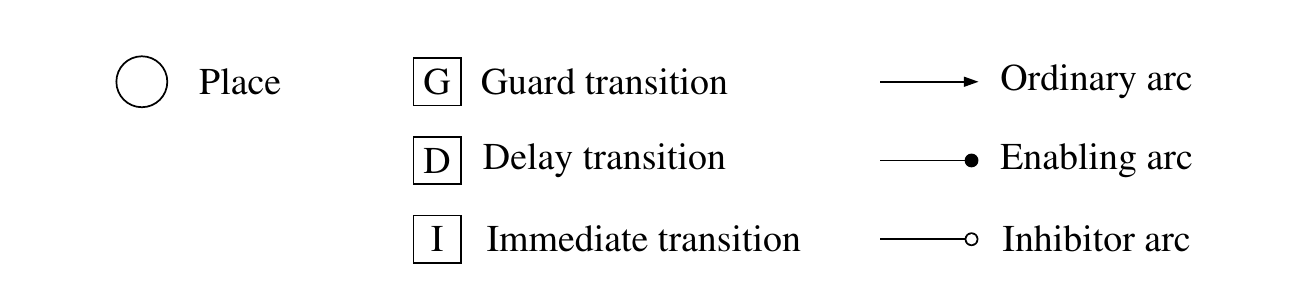}
\end{center}
\caption{Graphical notation for places, transitions and arcs in an SDCPN}
\label{fig:graphPTA}
\end{figure}

The SDCPN elements (${\cal P}$, ${\cal T}$, ${\cal A}$,
${\cal N}$, ${\cal S}$, ${\cal C}$, ${\cal I}$, ${\cal V}$, ${\cal
W}$, ${\cal G}$, ${\cal D}$, ${\cal F}$) are defined as follows:
\begin{itemize}
\item ${\cal P}$ is a finite set of places.

\item ${\cal T}$ is a finite set of transitions which consists
      of 1) a set ${\cal
      T}_G$ of guard transitions, 2) a set ${\cal T}_D$ of delay
      transitions and 3) a set ${\cal T}_I$ of immediate transitions.

\item ${\cal A}$ is a finite set of arcs which consists
      of 1) a set ${\cal A}_O$ of ordinary arcs, 2) a set ${\cal A}_E$ of
      enabling arcs and 3) a set ${\cal A}_I$ of inhibitor arcs.

\item ${\cal N} : {\cal A} \to {\cal P}\times {\cal T} \cup {\cal T} \times
      {\cal P}$ is a node function which maps each arc $A \in {\cal A}$ to
      a pair of ordered nodes ${\cal N}(A)$, where a node is a place or a
      transition.

\item ${\cal S} \subset \{\mathbb{R}^0, \mathbb{R}^1, \mathbb{R}^2, \ldots \}$
      is a finite set of colour types, with $\mathbb{R}^0 \triangleq \emptyset$.

\item ${\cal C} : {\cal P} \to {\cal S}$ is a colour type function which maps
      each place $P \in {\cal P}$ to a specific colour type.
      Each token in $P$ is to have a colour in ${\cal C}(P)$. If ${\cal C}(P)
      = \mathbb{R}^0$ then a token in $P$ has no colour.

\item ${\cal I}$ is a probability measure,
      which defines the initial marking of the net:
      for each place it defines a number $\geq 0$ of tokens
      initially in it and it defines their initial colours.

\item ${\cal V} = \{{\cal V}_P ; P \in {\cal P}, {\cal C}(P) \neq
      \mathbb{R}^0\}$ is a set of token colour functions. For each place
      $P\in {\cal P}$ for which ${\cal C}(P)\neq \mathbb{R}^0$, it
      contains a function ${\cal V}_P : {\cal C}(P)
      \to {\cal C}(P)$ that defines the drift coefficient of a
      differential equation for the colour of a token in place $P$.

\item ${\cal W} = \{{\cal W}_P ; P \in {\cal P}, {\cal C}(P)
      \neq \mathbb{R}^0\}$ is a set of token colour matrix functions.
      For each place $P\in {\cal P}$ for which ${\cal C}(P)\neq
      \mathbb{R}^0$, it contains a measurable mapping
      ${\cal W}_P : {\cal C}(P) \to \mathbb{R}^{n(P)\times h(P)}$
      that defines the diffusion coefficient of a stochastic
      differential equation for the colour of a token in place $P$,
      where $h : {\cal P} \rightarrow \mathbb{N}$ and $n : {\cal P}
      \rightarrow \mathbb{N}$ is such that ${\cal C}(P) = \mathbb{R}^{n(P)}$.
      It is assumed
      that ${\cal W}_P$ and ${\cal V}_P$ satisfy
      conditions that ensure a probabilistically unique solution of
      each stochastic differential equation.

\item ${\cal G}  = \{{\cal G}_T ; T \in {\cal T}_G\}$ is a set of
      transition guards. For each $T \in {\cal
      T}_G$, it contains a transition guard ${\cal G}_T$, which is an open
      Euclidean subset with boundary $\partial {\cal G}_T$.

\item ${\cal D} = \{{\cal D}_T ; T \in {\cal T}_D\}$ is a set of
      transition delay rates. For each $T \in {\cal T}_D$, it contains
      a locally integrable transition delay rate ${\cal D}_T$.

\item ${\cal F} = \{{\cal F}_T ; T \in {\cal T}\}$ is a set of firing
      measures. For each $T \in {\cal T}$,
      it contains a firing measure ${\cal F}_T$, which
      generates the number and colours of the tokens produced when
      transition $T$ fires, given the value of the vector
      that collects all input tokens:
      For each output arc, zero or one token is
      produced.
      For each fixed $H$, ${\cal F}_T(H;\cdot)$
      is measurable. For any $c$, ${\cal F}_T(\cdot;c)$ is a probability
      measure.

\end{itemize}
For the places, transitions and arcs, the graphical notation is as
in Figure \ref{fig:graphPTA}.

\subsection{SDCPN execution}
\label{subsec:SDCPN execution}

The execution of an SDCPN provides a series of increasing stopping
times, $\{ \tau_i ; i=0, 1, \ldots \}$, $\tau_0 =0$, with for $t \in
(\tau_k,\tau_{k+1})$ a fixed number of tokens per place and per
token a colour which is the solution of a stochastic differential
equation.

\paragraph{Initiation.}
The probability measure ${\cal I}$ characterises an initial
marking at $\tau_0$, i.e.\ it gives each place $P \in {\cal
P}$ zero or more tokens and gives each token in $P$ a colour in
${\cal C}(P)$, i.e.\ a Euclidean-valued vector.

\paragraph{Token colour evolution.}
For each token in each place $P$ for which ${\cal C}(P) \neq
\mathbb{R}^0$: if the colour of this token is equal to $C_0^P$ at
time $t=\tau_0$, and if this token is still in this place at time
$t > \tau_0$, then the colour $C_t^P$ of this token equals the
probabilistically unique solution of the stochastic differential
equation $dC_t^P= {\cal V}_P(C_t^P)dt +{\cal
W}_P(C_t^P)dB_t^{i,P}$ with initial condition $C_{\tau_0}^P =
C_0^P$, and with $\{B_t^{i,P}\}$ an $h(P)$-dimensional standard
Brownian motion. Each token in a place for which ${\cal C}(P)
= \mathbb{R}^0$ remains without colour.

\paragraph{Transition enabling.}
A transition $T$ is {\em pre-enabled} if it has at least one
token per incoming ordinary and enabling arc in each of its input
places and has no token in places to which it is connected by an
inhibitor arc. For each transition $T$ that is pre-enabled at
$\tau_0$, consider one token per ordinary and enabling arc in its
input places and write $C_t^T$, $t \geq \tau_0$, as the column vector
containing the colours of these tokens; $C_t^T$ evolves
through time according to its corresponding token colour functions.
If this vector is not unique
(i.e., if one input place contains several tokens per arc),
all possible such vectors are executed in parallel.
A transition $T$ is {\em enabled} if it
is pre-enabled and a second requirement holds true. For $T \in
{\cal T}_I$, the second requirement automatically holds true at
the time of pre-enabling. For $T \in {\cal T}_G$, the second
requirement holds true when $C_t^T \in \partial {\cal G}_T$.
For $T \in {\cal T}_D$, the second requirement holds true
at $t = \tau_0 + \sigma_1^T$, where $\sigma_1^T$ is
generated from a probability distribution function
$D_T(t-\tau_0) = 1-\mbox{exp}(-\int_{\tau_0}^t {\cal D}_{T}(C_s^{T})ds)$.
A Uniform random variable $U_i$ is used to determine this
$\sigma_1^T$.
In the case of competing enablings, the following
rules apply:
\begin{itemize}
\item[R0] The firing of an immediate transition has priority
over the firing of a guard or a delay transition.

\item[R1] If one transition becomes enabled by two or more sets
of input tokens at exactly the same time, and the firing of any
one set will not disable one or more other sets, then it will fire
these sets of tokens independently, at the same time.

\item[R2] If one transition becomes enabled by two or more sets
of input tokens at exactly the same time, and the firing of any
one set disables one or more other sets, then the set that is
fired is selected randomly, with the same probability for each
set.

\item[R3] If two or more transitions become enabled at exactly
the same time and the firing of any one transition will not
disable the other transitions, then they will fire at the same
time.

\item[R4] If two or more transitions become enabled at
exactly the same time and the firing of any one transition disables
some other transitions, then each combination of
transitions that can fire independently without leaving enabled
transitions gets the same probability of firing.

\end{itemize}

\paragraph{Transition firing.}
If $T$ is enabled, suppose this occurs at time $\tau_1$
and in a particular vector of token colours
$C_{\tau_1}^T$, it removes one token per ordinary input arc corresponding
with $C_{\tau_1}^T$ from each
of its input places (i.e.\ tokens are not removed along enabling arcs).
Next, $T$ produces zero or one token along each output arc: If
$(e_{\tau_1}^T,a_{\tau_1}^T)$ is a random hybrid
vector generated from probability measure ${\cal F}_T(\cdot; C_{\tau_1}^T)$
(by making use of a Uniform random variable $U_i$), then
vector $e_{\tau_1}^T$ is a vector of zeros and ones, where
the $i$th vector element corresponds with the $i$th outgoing
arc of transition $T$. An output place gets a token iff it is
connected to an arc that corresponds with a vector element 1.
Moreover, $a_{\tau_1}^T$ specifies the
colours of the produced tokens.

\paragraph{Execution from first transition firing onwards.}
At $t=\tau_1$, zero or more transitions are pre-enabled (if this
number is zero, no transitions will fire anymore). If these
include immediate transitions, then these are fired without delay,
but with use of rules R0--R4. If after this, still immediate
transitions are enabled, then these are also fired, and so forth,
until no more immediate transitions are enabled.
Next, the SDCPN is executed in the same way as described above for
the situation from $\tau_0$ onwards.

\subsection{SDCPN stochastic process}
\label{subsec:SDCPN process}

The marking of the SDCPN is given by the numbers of tokens in the
places and the associated colour values of these tokens and can be
mapped to a probabilistically unique SDCPN stochastic process
$\{M_t,C_t\}$ as follows: For any $t \geq \tau_0$, let a token
distribution be characterised by the vector $M_t' = (M_{1,t}',
\ldots, M_{|{\cal P}|,t}')$, where $M_{i,t}' \in \mathbb{N}$
denotes the number of tokens in place $P_i$ at time $t$ and $1,
\ldots ,|{\cal P}|$ refers to a unique ordering of places adopted
for SDCPN. At times $t \in (\tau_{k-1},\tau_k)$ when no transition
fires, the token distribution is unique and we define $M_t = M_t
'$. The associated colours of these tokens are gathered in a
column vector $C_t$ which first contains all colours of tokens in
place $P_1$, next (i.e.\ below it) all colours of tokens in place
$P_2$, etc, until place $P_{|{\cal P}|}$. If at time $t=\tau_k$
one or more transitions fire, then the SDCPN discrete process
state at time ${\tau_k}$ is defined by $M_{\tau_k} = $ the token
distribution that occurs after all transitions that fire at time
${\tau_k}$ have been fired. The associated colours of these tokens
are gathered in a column vector $C_{\tau_k}$ in the same way as
described above. This construction ensures that the process
$\{M_t, C_t\}$ has limits from the left and is continuous from the
right, i.e., it satisfies the c\`adl\`ag property.

\section{Air traffic example and its SDCPN model}\label{sec:airtrafficSDCPN}
To illustrate the advantages of SDCPN when modelling a complex
system, consider a simplified model of the evolution of an
aircraft in one sector of airspace. The deviation of this aircraft
from its intended path is affected by its engine system and its
navigation system. Each of these aircraft systems can be in either
{\em Working} (functioning properly) or {\em Not working}
(operating in some failure mode). Both systems switch between
these modes independently and with exponentially distributed
sojourn times, with finite rates $\delta_3$ (engine repaired),
$\delta_4$ (engine fails), $\delta_5$ (navigation repaired) and
$\delta_6$ (navigation fails), respectively. If both systems are
{\em Working}, the aircraft evolves in {\em Nominal} mode and the
position $Y_t$ and velocity $S_t$ of the aircraft are determined
by $dX_t = {\cal V}_1^{}(X_t)dt + {\cal W}_1 dW_t$, where $X_t =
(Y_t, S_t)'$. If either one, or both, of the systems is {\em Not
working}, the aircraft evolves in {\em Non-nominal} mode and the
position and velocity of the aircraft are determined by $dX_t =
{\cal V}_2^{}(X_t)dt + {\cal W}_2 dW_t$. The factors ${\cal
W}_1^{}$ and ${\cal W}_2^{}$ are determined by wind fluctuations.
Initially, the aircraft has position $Y_0$ and velocity $S_0$,
while both its systems are {\em Working}. The evaluation of this
process may be stopped when the aircraft has {\em Landed}, i.e.\
its vertical position and velocity are equal to zero.

\begin{figure}[h!bt]
\begin{center}
\includegraphics{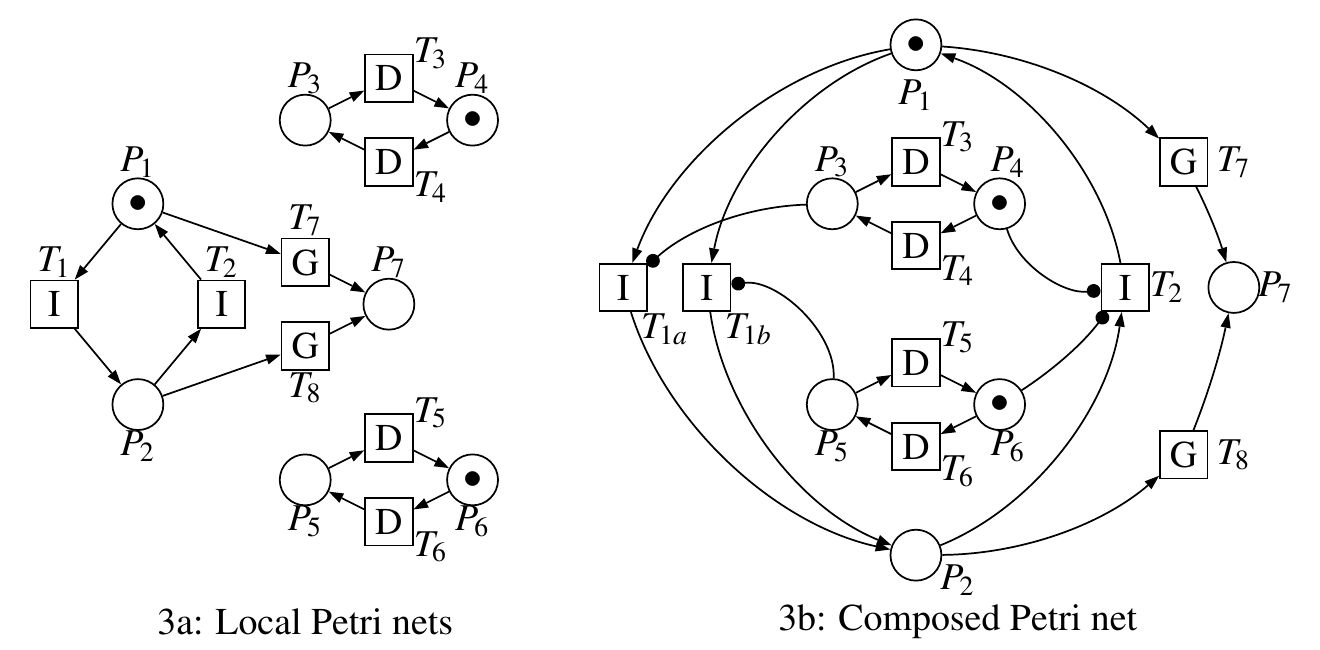}
\end{center}
\caption{SDCPN graph for the aircraft evolution example}
\label{fig:LPN example integrated}
\end{figure}

An SDCPN graph for this example is developed in two stages. In
the first stage, the agents of the operation are modelled separately,
by one local SDCPN each, see Fig. \ref{fig:LPN example integrated}a.
In the next stage, the interactions between
the agents are modelled, thus connecting the local SDCPN,
Fig. \ref{fig:LPN example integrated}b.

Fig.\ \ref{fig:LPN example integrated}b shows
the SDCPN graph for this example, where,
\begin{itemize}
\item $P_1$ denotes aircraft evolution {\em Nominal}, i.e.\
evolution is
      according to ${\cal V}_1$ and ${\cal W}_1$.
\item $P_2$ denotes aircraft evolution {\em Non-nominal}, i.e.\
evolution is
      according to ${\cal V}_2$ and ${\cal W}_2$.
\item $P_3$ and $P_4$ denote engine system {\em Not working} and
{\em
      Working}, respectively.
\item $P_5$ and $P_6$ denote navigation system {\em Not working}
and {\em
      Working}, respectively.
\item $P_7$ denotes the aircraft has landed. \item $T_{1a}$ and
$T_{1b}$ denote a transition of aircraft evolution from
      {\em Nominal} to {\em Non-nominal}, due to engine system or
      navigation system {\em Not working}, respectively.
\item $T_2$ denotes a transition of aircraft evolution from {\em
      Non-nominal} to {\em Nominal}, due to engine system and navigation
      system both {\em Working} again.
\item $T_3$ through $T_6$ denote transitions between {\em Working}
and {\em
      Not working} of the engine and navigation systems.
\item $T_7$ and $T_8$ denote transitions of the aircraft landing.

\end{itemize}

The graph in Fig.\ \ref{fig:LPN example integrated}b completely
defines SDCPN elements ${\cal P}$, ${\cal T}$, ${\cal A}$ and
${\cal N}$, where ${\cal T}_G = \{T_7, T_8\}$, ${\cal T}_D =
\{T_3, T_4, T_5, T_6\}$ and ${\cal T}_I = \{T_{1a}, T_{1b},
T_{2}\}$. The other SDCPN elements are specified below:
\begin{itemize}
\item[${\cal S}$:] Two colour types are defined; ${\cal S} =
\{\mathbb{R}^0, \mathbb{R}^6\}$.

\item[${\cal C}$:] ${\cal C}(P_1) = {\cal C}(P_2) = {\cal C}(P_7)
= \mathbb{R}^6$, i.e.\
                   tokens in $P_1$, $P_2$ and $P_7$ have colours in $\mathbb{R}^6$;
                   the colour components model the 3-dimensional position
                   and 3-dimensional velocity of the aircraft. ${\cal C}(P_3) = {\cal C}(P_4)
                   = {\cal C}(P_5) = {\cal C}(P_6) =\mathbb{R}^0 \triangleq \emptyset$.

\item[${\cal I}$:] Place $P_1$ initially has a token with colour
$X_0 =
                   (Y_0, S_0)'$, with $Y_0 \in \mathbb{R}^2 \times (0,\infty)$
                   and $S_0 \in \mathbb{R}^3 \setminus \mbox{Col}\{0,0,0\}$.
                   Places $P_4$ and $P_6$ initially each have a token without
                   colour.

\item[${\cal V}$,] ${\cal W}$: The token colour functions for
places $P_1$, $P_2$ and
                   $P_7$ are determined by
                   $({\cal V}_{1}^{}, {\cal W}_1^{})$,
                   $({\cal V}_{2}^{}, {\cal W}_2^{})$, and
                   $({\cal V}_{7}^{}, {\cal W}_7^{})$, respectively, where
                   $({\cal V}_{7}^{}, {\cal W}_7^{})=(0,0)$.
                   For places $P_3$ -- $P_6$ there is no token colour function.

\item[${\cal G}$:] Transitions $T_7$ and $T_8$ have a guard
                   defined by ${\cal G}_{T_7} = {\cal G}_{T_8} =
                   \mathbb{R}^2 \times (0, \infty) \times \mathbb{R}^2
                   \times (0, \infty)$.

\item[${\cal D}$:] The jump rates for transitions $T_3$, $T_4$,
$T_5$ and
                   $T_6$ are ${\cal D}_{T_3}(\cdot) = \delta_3$,
                   ${\cal D}_{T_4}(\cdot) = \delta_4$, ${\cal D}_{T_5}(\cdot) =
                   \delta_5$ and ${\cal D}_{T_6}(\cdot) = \delta_6$.

\item[${\cal F}$:] Each transition has a unique output place, to
which it
                   fires a token with a colour (if applicable) equal to the
                   colour of the token removed.

\end{itemize}

\section{From SDCPN to GSHS}\label{sec:gshsautomaton}

Following \cite{BujorianuLygeros2006}, this section first presents a
definition of \emph{general stochastic hybrid system} (GSHS) and
its execution. In \cite{EverdijBlom2006} it has been proven that
under a few conditions, SDCPN and GSHS are bisimilar. In
Subsection \ref{sec:sdcpnexample} this is illustrated
by showing how the SDCPN example of the previous section can be
mapped to a bisimilar GSHS.

\subsection{GSHS definition}
\paragraph{Definition 4.1} (General stochastic hybrid system)\label{def:gshs}
\emph{A GSHS is an automaton (${\bf K}$, $d$, ${\cal X}$,
$f$, $g$, \emph{Init}, $\lambda$, $Q$), where
\begin{itemize}
\item ${\bf K}$ is a countable set.
\item $d: {\bf K} \rightarrow \mathbb{N}$ maps each $\theta \in {\bf K}$
      to a natural number.
\item ${\cal X}: {\bf K} \rightarrow \{E_{\theta} ; \theta \in {\bf K}\}$
      maps each $\theta \in {\bf K}$ to an open subset $E_{\theta}$ of $\mathbb{R}^{d(\theta)}$.
      With this, the hybrid state space is given by $E \triangleq \{\{\theta\}
      \times E_{\theta}; \theta \in {\bf K}\}$.
\item $f: E \to \{\mathbb{R}^{d(\theta)} ; \theta \in {\bf
      K}\}$ is a vector field.
\item $g: E \to \{\mathbb{R}^{d(\theta)\times h} ; \theta \in {\bf K}\}$ is a
      matrix field, with $h \in \mathbb{N}$.
\item \emph{Init}$: {\cal B}(E) \rightarrow [0,1]$ is an initial probability
      measure, with ${\cal B}(E)$ the Borel $\sigma$-algebra on $E$.
\item $\lambda: E \to \mathbb{R}^{+}$ is a jump rate function.
\item $Q : \mathcal{B}(E) \times (E \cup \partial E) \rightarrow [0,1]$ is a
      GSHS transition measure, where $\partial E \triangleq \{\{\theta\}
      \times \partial E_{\theta}; \theta \in {\bf K}\}$ is the boundary of
      $E$, in which $\partial E_{\theta}$ is the boundary of $E_{\theta}$.
\end{itemize}
}

\paragraph{Definition 4.2} (GSHS execution)\label{def:GSHSexecution}
\emph{A stochastic process $\{\theta_t,X_t\}$ is called a \emph{GSHS
execution} if there exists a sequence of stopping times $0=\tau_0
< \tau_1 < \tau_2 \cdots$ such that for each $k \in \mathbb{N}$:
\begin{itemize}
\item $(\theta_0,X_0)$ is an $E$-valued random variable extracted
      according to probability measure \emph{Init}.
\item For $t \in [\tau_k,\tau_{k+1})$, $\theta_t=\theta_{\tau_k}$
      and $X_t=X_t^k$, where for $t\geq \tau_k$, $X_t^k$ is a
      solution of the stochastic differential equation $dX_t^k =
      f(\theta_{\tau_k},X_t^k)dt + g(\theta_{\tau_k},X_t^k)dB_t^{\theta_{\tau_k}}$
      with initial condition $X_{\tau_k}^k=X_{\tau_k}$, and where
      $\{B_t^{\theta}\}$ is $h$-dimensional standard Brownian motion for
      each $\theta \in {\bf K}$.
\item $\tau_{k+1} = \tau_k + \sigma_k$, where $\sigma_k$ is chosen
      according to a survivor function given by $F(t) =$ \\
      ${\bf 1}_{(t < \tau^*)} \exp ( -\int_0^t \lambda(\theta,X_s^k)ds)$.
      Here, $\tau^* = \inf \{t > \tau_k \mid X_t^k \in \partial E_{\theta_{\tau_k}}\}$
      and ${\bf 1}$ is indicator function.
\item The probability distribution of $(\theta_{\tau_{k+1}}, X_{\tau_{k+1}})$, i.e.\ the
      hybrid state right after the jump, is governed by the law $Q(\cdot;
      (\theta_{\tau_k},X_{\tau_{k+1}-}))$.
\end{itemize}}

\cite{BujorianuLygeros2006} show that under assumptions G1-G4 below,
a GSHS execution is a strong Markov Process and has the c{\`a}dl{\`a}g property
(right continuous with left hand limits).
\begin{itemize}
\item[G1] $f(\theta, \cdot)$ and $g(\theta, \cdot)$ are Lipschitz continuous and bounded.
      This yields that for each initial state $(\theta,x)$ at initial time $\tau$
      there exists a pathwise unique solution $X_t$ to
      $dX_t = f(\theta,X_t)dt + g(\theta, X_t)dB_t$, where
      $\{B_t\}$ is $h$-dimensional standard Brownian motion.

\item[G2] $\lambda: E \to \mathbb{R}^{+}$
      is a measurable function such that for all $\xi \in E$, there is $\epsilon(\xi) > 0$ such that
      $t \to \lambda(\theta_t,X_t)$ is integrable on $[0,\epsilon(\xi))$.

\item[G3] For each fixed $A \in {\cal B}(E)$, the map $\xi \to
Q(A;\xi)$ is measurable and
      for any $(\theta,x) \in E \cup \partial E$, $Q(\cdot; \theta,x)$ is a probability measure.

\item[G4] If $N_t = \sum_k {\bf 1}_{(t \geq \tau_k)}$, then it is assumed that for every starting point $(\theta,x)$ and
      for all $t \in \mathbb{R}^{+}$, $\mathbb{E}N_t < \infty$. This means, there will be a finite
      number of jumps in finite time.
\end{itemize}

\subsection{A bisimilar GSHS for the example SDCPN}\label{sec:sdcpnexample}
Next we transform the SDCPN example model of Section \ref{sec:airtrafficSDCPN}
into a bisimilar GSHS. The first step is to construct the state space
${\bf K}$ for the GSHS discrete process $\{\theta_t\}$. This is
done by identifying the  SDCPN {\em reachability graph}. Nodes in
the reachability graph provide the number of tokens in each of the
SDCPN places. Arrows connect these nodes as they represent
transitions firing. The SDCPN of Fig.\ \ref{fig:LPN example
integrated}b has seven places hence the reachability graph for this
example has elements that are vectors of length 7. These nodes,
excluding the nodes that enable immediate transitions, form the
GSHS discrete state space.

\begin{figure}[h!bt]
\begin{center}
\includegraphics{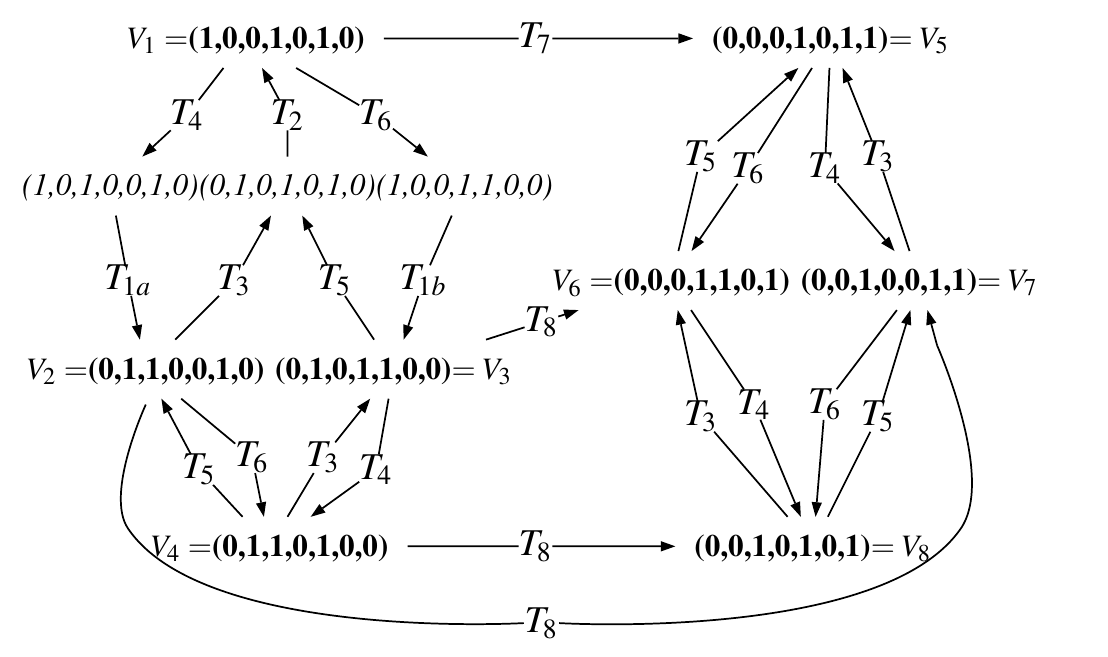}
\end{center}
\caption{Reachability graph for the SDCPN of Fig.\ 3b.
The nodes in bold type face correspond with
the elements of the GSHS discrete state space ${\bf K}$.}
\label{fig:RGexample}
\end{figure}

The reachability graph is shown in Fig.\ \ref{fig:RGexample}, with
nodes that form the GSHS discrete state space in Bold typeface,
i.e.\ ${\bf K} = \{V_1, \ldots, V_8\}$, with
$V_1=(1,0,0,1,0,1,0)$, $V_2=(0,1,1,0,0,1,0)$,
$V_3=(0,1,1,0,1,0,0)$, $V_4=(0,1,0,1,1,0,0)$,
$V_5=(0,0,0,1,0,1,1)$, $V_6=(0,0,1,0,0,1,1)$,
$V_7=(0,0,1,0,1,0,1)$, $V_8=(0,0,0,$ $1,1,0,1)$. Since initially
there is a token in places $P_1$, $P_4$ and $P_6$, the GSHS
initial mode equals $\theta_0 = V_1 = (1,0,0,1,0,1,0)$. The GSHS
initial continuous state value equals the vector containing the
initial colours of all initial tokens. Since the initial colour of
the token in Place $P_1$ equals $X_0$, and the tokens in places
$P_4$ and $P_6$ have no colour, the GSHS initial continuous state
value equals Col$\{X_0, \emptyset, \emptyset\} = X_0$. The GSHS
drift coefficient $f$ is given by $f(\theta,\cdot)={\cal
V}_1(\cdot)$ for $\theta=V_1$, $f(\theta,\cdot) ={\cal
V}_2(\cdot)$ for $\theta \in \{V_2, V_3, V_4\}$, and
$f(\theta,\cdot)=0$ otherwise. For the diffusion coefficient,
$g(\theta,\cdot)={\cal W}_1$ for $\theta=V_1$,
$g(\theta,\cdot)={\cal W}_2$ for $\theta \in \{V_2, V_3, V_4\}$,
and $g(\theta,\cdot)=0$ otherwise. The hybrid state space is given
by $E = \{\{\theta\} \times E_{\theta} ; \theta \in \mathbb{M}\}$,
where for $\theta \in \{ V_1, V_2, V_3, V_4\}$: $E_{\theta} =
\mathbb{R}^2 \times (0,\infty) \times \mathbb{R}^2 \times
(0,\infty)$ and for $\theta \in \{ V_5, V_6, V_7, V_8\}$:
$E_{\theta} = \mathbb{R}^6$. Always two delay transitions are
pre-enabled: either $T_3$ or $T_4$ and either $T_5$ or $T_6$. This
yields $\lambda(V_1, \cdot) = \lambda(V_5, \cdot) = \delta_4 +
\delta_6$, $\lambda(V_2, \cdot) = \lambda(V_6, \cdot) = \delta_3 +
\delta_6$, $\lambda(V_3, \cdot) = \lambda(V_7, \cdot) = \delta_3 +
\delta_5$, $\lambda(V_4, \cdot) = \lambda(V_8, \cdot) = \delta_4 +
\delta_5$. For the determination of GSHS transition measure $Q$, we make
use of the reachability graph, the sets ${\cal D}$, ${\cal G}$ and
${\cal F}$ and the rules R0--R4. In Table \ref{tab:Q},
$Q(\theta',x'; \theta,x)=p$ denotes that if $(\theta,x)$ is the
value of the GSHS state before the hybrid jump, then, with
probability $p$, $(\theta',x')$ is the value of the GSHS state
immediately after the jump.

\begin{table}[ht]
\centering \caption{Example GSHS transition measure for size of jump}
\label{tab:Q}
\begin{tabular}{|lll|}
\hline
For $x \notin \partial E_{V_1}$:  &
   $Q(V_2,x;V_1,x) = \frac{\delta_4}{\delta_4+\delta_6}$,    &
   $Q(V_4,x;V_1,x) = \frac{\delta_6}{\delta_4+\delta_6}$ \\
For $x \in \partial E_{V_1}$:    & $Q(V_5,x;V_1,x) = 1$ &  \\
\hline For $x \notin \partial E_{V_2}$:    &
   $Q(V_3,x;V_2,x) = \frac{\delta_6}{\delta_3+\delta_6}$,   &
   $Q(V_1,x;V_2,x) = \frac{\delta_3}{\delta_3+\delta_6}$ \\
For $x \in \partial E_{V_2}$:    & $Q(V_6,x;V_2,x) = 1$    &  \\
\hline For $x \notin \partial E_{V_3}$:    &
   $Q(V_4,x;V_3,x) = \frac{\delta_3}{\delta_3+\delta_5}$,    &
   $Q(V_2,x;V_3,x) = \frac{\delta_5}{\delta_3+\delta_5}$ \\
For $x \in \partial E_{V_3}$:    & $Q(V_7,x;V_3,x) = 1$    &  \\
\hline For $x \notin \partial E_{V_4}$:    &
   $Q(V_3,x;V_4,x) = \frac{\delta_4}{\delta_4+\delta_5}$,   &
   $Q(V_1,x;V_4,x) = \frac{\delta_5}{\delta_4+\delta_5}$ \\
For $x \in \partial E_{V_4}$:    & $Q(V_8,x;V_4,x) = 1$    &  \\
\hline For all $x$:  & $Q(V_6,x;V_5,x) =
\frac{\delta_4}{\delta_4+\delta_6}$,    & $Q(V_8,x;V_5,x) =
\frac{\delta_6}{\delta_4+\delta_6}$ \\ \hline For all $x$:    &
$Q(V_7,x;V_6,x) = \frac{\delta_6}{\delta_3+\delta_6}$,    &
$Q(V_5,x;V_6,x) = \frac{\delta_3}{\delta_3+\delta_6}$ \\ \hline
For all $x$:    & $Q(V_8,x;V_7,x) =
\frac{\delta_3}{\delta_3+\delta_5}$,    & $Q(V_6,x;V_7,x) =
\frac{\delta_5}{\delta_3+\delta_5}$ \\ \hline For all $x$:    &
$Q(V_7,x;V_8,x) = \frac{\delta_4}{\delta_4+\delta_5}$,   &
$Q(V_5,x;V_8,x) =
\frac{\delta_5}{\delta_4+\delta_5}$ \\
\hline
\end{tabular}
\end{table}

With this, the SDCPN of the aircraft evolution example is uniquely
mapped to an GSHS. It can be shown that the SDCPN execution and the
execution of the resulting GSHS are probabilistically equivalent,
i.e.\ the SDCPN and the GSHS are bisimilar. Thanks to this bisimilarity
we can now use the automata framework to analyse the GSHP that is
defined by the SDCPN model for the example.

\section{From SDCPN to HSDE}\label{sec:hsde}

Following \cite{Blom2003} and
\cite{BlomBakkerEverdijPark2003b}, this section first presents
a definition of \emph{hybrid stochastic differential equation} (HSDE) and gives conditions
under which the HSDE has a pathwise unique solution. This pathwise
unique solution is referred to as \emph{HSDE solution process} or
GSHP. The basic advantage of using HSDE in defining a GSHP over
using GSHS is that with the HSDE approach the spontaneous jump
mechanism is explicitly built on an underlying stochastic basis,
whereas in GSHS the execution itself imposes an underlying
stochastic basis.
In \cite{EverdijBlom2009} it has been proven that
under a few conditions, SDCPN and HSDE are bisimilar. In
Subsection \ref{sec:sdcpnexample2} this is illustrated
by showing how the SDCPN example of the previous section can be
mapped to a bisimilar HSDE.


\subsection{HSDE definition}

For the HSDE setting we start with a complete stochastic basis
$(\Omega, \Im, \mathbb{F}, \mathbb{P}, \mathbb{T})$, in which a
complete probability space $(\Omega, \Im, \mathbb{P})$ is equipped
with a right-continuous filtration $\mathbb{F} = \{\Im_t\}$ on the
positive time line $\mathbb{T}=\mathbb{R}^{+}$. This stochastic
basis is endowed with a probability measure $\mu_{\theta_0, X_0}$
for the initial state,
an independent $h$-dimensional standard Wiener process $\{W_t\}$ and an
independent homogeneous Poisson random measure $p_P(dt,dz)$ on
$\mathbb{T} \times \mathbb{R}^{d+1}$.

\paragraph{Definition 5.1} (Hybrid stochastic differential equation)\label{def:hsde}
\emph{An HSDE on stochastic
basis $(\Omega, \Im, \mathbb{F}, \mathbb{P}, \mathbb{T})$, is
defined as a set of equations (\ref{eq:xister4a})-(\ref{eq:Q4}) in
which a collection of elements ($\mathbb{M}$, $E$, $f$, $g$,
$\mu_{\theta_0,X_0}$, $\Lambda$, $\psi$, $\rho$, $\mu$, $p_P$,
$\{W_t\}$) appear.}\\[5mm]
The elements ($\mathbb{M}$, $E$, $f$, $g$, $\mu_{\theta_0,X_0}$,
$\Lambda$, $\psi$, $\rho$, $\mu$, $p_P$, $\{W_t\}$) are defined as
follows:
\begin{itemize}
\item $\mathbb{M} = \{\vartheta_1, \ldots ,  \vartheta_N\}$ is a
      finite set, $N \in \mathbb{N}$, $1 \leq N < \infty$.
\item $E = \{\{\theta\} \times E_{\theta} ; \theta \in
      \mathbb{M}\}$ is the hybrid state space,
      where for each $\theta \in \mathbb{M}$, $E_{\theta}$ is
      an open subset of $\mathbb{R}^{n}$ with boundary
      $\partial E_{\theta}$. The boundary of $E$ is
      $\partial E = \{\{\theta\} \times \partial E_{\theta};
      \theta \in \mathbb{M}\}$.
\item $f: \mathbb{M} \times \mathbb{R}^{n} \to \mathbb{R}^{n}$
      is a measurable mapping.
\item $g: \mathbb{M} \times \mathbb{R}^{n} \to
      \mathbb{R}^{n\times h}$ is a  measurable mapping.
\item $\mu_{\theta_0, X_0}: \Omega \times {\cal B}(E) \to [0,1]$
is a probability
      measure for the initial random variables
      $\theta_0$, $X_0$, which are defined on the stochastic
      basis; $\mu_{\theta_0, X_0}$ is assumed to be invertible.
\item $\Lambda: \mathbb{M} \times \mathbb{R}^{n} \to [0,\infty)$
      is a measurable mapping.
\item $\psi: \mathbb{M} \times \mathbb{M} \times \mathbb{R}^{n}
      \times \mathbb{R}^{d} \to \mathbb{R}^{n}$ is a measurable mapping
      such that $x  + \psi(\vartheta, \theta, x,\underline{z}) \in
      E_{\vartheta}$ for all $x \in E_{\theta}$, $\underline{z}\in
      \mathbb{R}^d$, and $\vartheta, \theta\in\mathbb{M}$.
\item $\rho: \mathbb{M} \times \mathbb{M} \times \mathbb{R}^{n}
      \to [0,\infty)$ is a measurable mapping such that
      $\sum_{i=1}^N \rho(\vartheta_i, \theta, x) =1$ for all $\theta \in
      \mathbb{M}, x \in \mathbb{R}^{n}$.
\item $\mu: \Omega \times \mathbb{R}^d \to [0,1]$ is a probability
      measure which is assumed to be invertible.
\item $p_P : \Omega \times \mathbb{T} \times \mathbb{R}^{d+1} \to \{0,1\}$
      is a homogeneous Poisson random measure on the stochastic basis,
      independent of $(\theta_0,X_0)$. The intensity measure
      of $p_P(dt,dz)$ equals $dt \cdot \mu_L(dz_1) \cdot \mu(d\underline{z})$, where
      $z = \mbox{Col}\{z_1, \underline{z}\}$ and $\mu_L$ is the Lebesgue measure.
\item $W : \Omega \times \mathbb{T} \to \mathbb{R}^h$ such that $\{W_t\}$ is
      an $h$-dimensional standard Wiener process on the stochastic basis, and
      independent of $(\theta_0,X_0)$ and $p_P$.
\end{itemize}
Using these elements, the HSDE process $\{\theta_t^*, X_t^*\}$ is
defined as follows:
\vspace{-1mm}
\begin{equation}\label{eq:xister4a}
\theta_t^* = \theta_t^k  \mbox{ for all } t \in [\tau_k^b,
\tau_{k+1}^b), k=0, 1, 2, \ldots
\end{equation}
\begin{equation}\label{eq:xister4b}
X_t^* = X_t^k  \mbox{ for all } t \in [\tau_k^b, \tau_{k+1}^b),
k=0, 1, 2, \ldots
\end{equation}
Hence $\{\theta_t^*, X_t^*\}$ consists of a concatenation of
processes $\{\theta_t^k, X_t^k\}$ which are defined by
(\ref{eq:dthetat4})-(\ref{eq:Q4}) below. If the system
(\ref{eq:xister4a})-(\ref{eq:Q4}) has a solution in probabilistic
sense, then the process $\{\theta_t^*,
X_t^*\}$ is referred to as \emph{HSDE solution process}
or \emph{GSHP}.
\vspace{-3mm}
\begin{equation}\label{eq:dthetat4}
d\theta_{t}^k = \sum_{i=1}^N (\vartheta_i -  \theta_{t-}^k)
p_P(dt, (\Sigma_{i-1}(\theta_{t-}^k,X_{t-}^k),
\Sigma_{i}(\theta_{t-}^k, X_{t-}^k)] \times \mathbb{R}^d)
\end{equation}
\vspace{-3mm}
\begin{equation}\label{eq:dxt4}
dX_t^k = f(\theta_t^k, X_t^k)dt + g(\theta_t^k, X_t^k)dW_t +
\int_{\mathbb{R}^d} \psi(\theta_t^k, \theta_{t-}^k, X_{t-}^k,
\underline{z}) p_P(dt, (0, \Lambda(\theta_{t-}^k,X_{t-}^k)] \times
d\underline{z})
\end{equation}
with $\theta_0^0 = \theta_0$, $X_0^0 = X_0$ and with $\Sigma_0$
through $\Sigma_N$ measurable mappings satisfying, for $\theta \in
\mathbb{M}$, $\vartheta_j \in \mathbb{M}$, $x \in \mathbb{R}^{n}$:
\vspace{-2mm}
\begin{equation}\label{eq:sigma4}
\Sigma_i(\theta,x) = \left\{
\begin{array}{ll}
\Lambda(\theta,x) \sum_{j=1}^i \rho(\vartheta_j, \theta,x) & \mbox{if } i>0  \\
0   & \mbox{if } i=0
\end{array} \right.
\end{equation}
In addition, for $k = 0, 1, 2, \ldots$, with $\tau_0^b = 0$:
\vspace{-2mm}
\begin{equation}\label{eq:taui4}
\tau_{k+1}^b \triangleq \inf \{t > \tau_{k}^b \mid (\theta_t^{k},
X_t^{k})\in \partial E\}
\end{equation}
\begin{equation}\label{eq:pxii4}
\mathbb{P}\{\theta_{\tau_{k+1}^b}^{k+1} = \vartheta,
X_{\tau_{k+1}^b}^{k+1} \in {A} \mid \theta_{\tau_{k+1}^b-}^{k}
=\theta, X_{\tau_{k+1}^b-}^{k} = x \} = Q(\{\vartheta\} \times A;
\theta,x)
\end{equation}
for $A \in {\cal B}(\mathbb{R}^n)$, where $Q$ is given by
\vspace{-2mm}
\begin{equation}\label{eq:Q4}
Q(\{\vartheta\}\times A; \theta, x) = \rho(\vartheta,\theta, x)
\int_{\mathbb{R}^d} {\bf 1}_{A} (x + \psi (\vartheta, \theta, x,
\underline{z})) \mu(d\underline{z})
\end{equation}

Next, the following proposition can be shown to hold true
\cite{EverdijBlom2009}:

\paragraph{Proposition 5.1}\label{th:BlomThirdTh}
\emph{Let conditions H1-H8 below hold true. Let $(\theta_0^* (\omega),
X_0^*(\omega)) = (\theta_0, X_0)\in E$ for all $\omega$. Then for
every initial condition $(\theta_0, X_0)$,
(\ref{eq:xister4a})-(\ref{eq:Q4}) has a pathwise unique solution
$\{\theta_t^*, X_t^*\}$ which is c\`adl\`ag and adapted and is a
semi-martingale assuming values in the hybrid state space $E$.}

\begin{description}
\item[H1] For all $\theta \in \mathbb{M}$ there exists a constant
$K(\theta)$ such that for all $x \in \mathbb{R}^{n}$, $|
f(\theta,x) |^2 + \| g(\theta,x)) \|^2 \leq K(\theta)(1+ |x|^2)$,
where $|a|^2 = \sum_i(a_i)^2$ and $||b||^2 =
\sum_{i,j}(b_{ij})^2$.

\item[H2] For all $r \in \mathbb{N}$ and for all $\theta \in
\mathbb{M}$ there exists a constant $L_r(\theta)$ such that for
all $x$ and $y$ in the ball $B_r = \{ z \in \mathbb{R}^{n} \mid
|z| \leq r+1\}$, $| f(\theta,x) - f(\theta,y) |^2 + \| g(\theta,x)
-g(\theta,y) \|^2 \leq L_r(\theta) |x-y|^2$.

\item[H3] For each $\theta \in \mathbb{M}$, the mapping $\Lambda(\theta,
\cdot): \mathbb{R}^{n} \to [0,\infty)$ is continuous and
bounded, with upper bound a constant $C_{\Lambda}$.

\item[H4] For all $(\theta, \vartheta) \in \mathbb{M}^2$, the mapping
$\rho(\vartheta, \theta,\cdot): \mathbb{R}^{n} \to [0,\infty)$
is continuous.

\item[H5] For all $r \in \mathbb{N}$ there exists a constant
$M_r(\theta)$ such that
\vspace{-3mm}
\[
\sup_{|x| \leq r} \int_{\mathbb{R}^d} | \psi(\vartheta, \theta,x,
\underline{z}) | \mu(d\underline{z}) \leq M_r (\theta), \mbox{ for
all } \vartheta, \theta \in \mathbb{M}
\]
\item[H6] $| \psi(\theta, \theta, x ,
\underline{z})| = 0$ or $>1$ for all $\theta \in \mathbb{M}$, $x
\in \mathbb{R}^{n}$, $\underline{z} \in \mathbb{R}^d$

\item[H7] $\{(\theta_t^*,X_t^*)\}$ hits the
boundary $\partial E$ a finite number of times on
any finite time interval

\item[H8] $|\vartheta_i - \vartheta_j|>1$ for $i \neq j$, with
$| \cdot |$ a suitable metric well defined on $\mathbb{M}$.

\end{description}

\subsection{A bisimilar HSDE for the example SDCPN}\label{sec:sdcpnexample2}
Next we transform the SDCPN example model of Section \ref{sec:airtrafficSDCPN}
into a bisimilar HSDE. This mapping follows much the same procedure
as for SDCPN to GSHS,
except that the discrete state space is now referred to as $\mathbb{M}$
(rather than ${\bf K}$) and the Markov jump rate is now referred to as
$\Lambda$ (rather than $\lambda$). The main additional difference
is that the HSDE elements do not include a transition measure $Q$ to
define the size of jump, but include functions $\psi$, $\rho$ and
$\mu$ instead. The mapping of SDCPN to HSDE uses the construction of
transition measure $Q$ as an intermediate step, however.
For the particular example SDCPN in this paper, these
functions can be determined from $Q$ as follows:
Since the continuous valued process
jumps to the same value with probability 1, we find that
$\psi(V^i, V^j, x, \underline{z}) = 0$ for all $V^i$, $V^j$, $x$,
$\underline{z}$. Moreover, $\rho(V^i, V^j, x) = P_Q(V^i, x, V^j,
x)$ and $\mu$ may be any given invertible probability measure.

With this, the SDCPN of the aircraft evolution example is uniquely
mapped to an HSDE. If in addition, we want to make use of the HSDE
properties of Proposition 5.1, i.e.\ the
resulting HSDE solution process being adapted and a
semi-martingale, we need to make sure that HSDE conditions H1-H8
are satisfied. It is shown below that they are, under the
following sufficient condition D1 for the example SDCPN.
\begin{description}
\item[D1] For $P\in \{P_1, P_2\}$, there exist $K_P^v$, $L_P^v$,
$K_P^w$
and $L_P^w$ such that for all $c, a \in {\cal C}(P)$,\\
$|{\cal V}_P(c)|^2 \leq K_P^v(1+|c|^2)$ and
$|{\cal V}_P(c) - {\cal V}_P(a)|^2 \leq L_P^v|c-a|^2$ and\\
$\|{\cal W}_P(c)\|^2 \leq K_P^w(1+|c|^2)$ and $\|{\cal W}_P(c) -
{\cal W}_P(a)\|^2 \leq L_P^w|c-a|^2$.
\end{description}
We verify that under condition D1, HSDE conditions H1-H8 hold true
in this example.
\begin{description}
\item[H1:] From the construction of $f$ and $g$ above we have for
$\theta = V_1$: $|f(\theta,x)|^2 + \|g(\theta,x)\|^2 = |{\cal
V}_{1}(x)|^2 + \|{\cal W}_{1}(x) \|^2 \leq  K_{P_1}^v (1+|x|^2) +
K_{P_1}^w (1+|x|^2) =  K(\theta)(1+|x|^2)$, with $K(\theta) =
(K_{P_1}^v+K_{P_1}^w)$. For $\theta = V_2, V_3, V_4$ the
verification is with replacing ${\cal V}_{1}$, ${\cal W}_{1}$ by
${\cal V}_{2}$, ${\cal W}_{2}$.

\item[H2:] From the construction of $f$ and $g$ above we have for
$\theta = V_1$: $| f(\theta,x) - f(\theta,y) |^2 + \| g(\theta,x)
-g(\theta,y) \|^2 = | {\cal V}_{1}(x) - {\cal V}_{1}(y)|^2
  +  \| {\cal W}_{1}(x) - {\cal W}_{1}(y)\|^2
\leq L_{P_1}^v |x-y|^2
 +  L_{P_1}^w |x-y|^2
 =  L_r (\theta) |x-y|^2$
with $L_r (\theta) = L_{P_1}^v+L_{P_1}^w$. For $\theta = V_2, V_3,
V_4$ replace ${\cal V}_{1}$, ${\cal W}_{1}$ by ${\cal V}_{2}$,
${\cal W}_{2}$.

\item[H3:] Since $\delta_3$--$\delta_6$ are constant, for all
$\theta$, $\Lambda(\theta, \cdot)$ is bounded and continuous, with
upper bound $C_{\Lambda} = \max \{ \delta_4 + \delta_6, \delta_3 +
\delta_6, \delta_3 + \delta_5, \delta_4 + \delta_5\}$.

\item[H4:] Since for all $\theta, \vartheta$, $P_Q(\vartheta,
\cdot; \theta, x)$ is constant, we find $\rho(\vartheta, \theta,
x) = P_Q(\vartheta, x, \theta, x)$ is continuous.

\item[H5 and H6:] These are satisfied due to $\psi(V^i, V^j, x,
\underline{z}) = 0$ for all $V^i$, $V^j$, $x$, $\underline{z}$.

\item[H7:] This condition holds due to $\delta_3$--$\delta_6$
being finite and the fact that in this SDCPN example, there is no
firing sequence of more than one guard transition. \item[H8:] This
condition holds for all $V_1, \ldots, V_8$, with metric $|a|^2 =
\sum_i(a_i)^2$.
\end{description}
Thanks to this bisimilarity mapping we can now use HSDE tools to
analyse the GSHP that is defined by the execution of the SDCPN
model for the example.

\section{Conclusions}\label{sec:conclusions}
The aim of this paper was to explain bisimilarity relations
between SDCPN (stochastically and dynamically coloured Petri net),
GSHS (general stochastic hybrid system) and HSDE (hybrid
stochastic differential equation), which means that the strengths
of one stochastic model formalism can be used by both of the other
stochastic model formalisms. More specifically, these bisimilarity
relations make it possible to combine the formal verification
power of automata with the analysis power of stochastic
differential equations and the compositional specification power
of Petri nets.

We started in Section \ref{sec:sdcpn} by defining SDCPN and the
resulting SDCPN stochastic process, which is referred to as a GSHP
(general stochastic hybrid process). In Section
\ref{sec:airtrafficSDCPN} we presented a simple but illustrative
SDCPN example model. In Section \ref{sec:gshsautomaton} we studied
GSHP as an execution of a GSHS and illustrated by using the
example of Section \ref{sec:airtrafficSDCPN} that SDCPN and GSHS
are bisimilar. Next, in Section \ref{sec:hsde} we studied GSHP as
a stochastic process solution of HSDE and showed with an
illustrative example that SDCPN and HSDE are bisimilar.

The bisimilarities between SDCPN, GSHS and HSDE models for the
example considered mean that the resulting example model inherits
the strengths of all three formal stochastic modelling formalisms.
This has been depicted in Fig.\ \ref{fig:GSHS:GSHP:SDCPN} in the
introduction. Examples of GSHP properties are convergence in
discretisation, existence of limits, existence of event
probabilities, strong Markov properties, reachability analysis.
Examples of GSHS features are their connection to formal methods
in automata theory and optimal control theory. Examples of HSDE
features are stochastic analysis tools for semi-martingales.
Examples of SDCPN features are natural expression of causal
dependencies, concurrency and synchronisation mechanism,
hierarchical and modular construction, and graphical
representation. These complementary advantages of SDCPN, GSHS,
HSDE and GSHP perspectives tend to increase with the complexity of
the system considered.

An illustrative large scale application of bisimularity relations
between SDCPN, HSDE and stochastic hybrid automata has been
developed in air traffic management. Currently pilots depend of
air traffic controllers in solving potential conflicts between
their flight trajectories. This places a huge requirement on the
tasks of an air traffic controller. Imagine a similar kind of
approach for road traffic; then each car driver would be blind and
depends of instructions that some road traffic controller is
communicating with the car drivers. How many cars do you think can
be managed by one road traffic controller? The number of aircraft
that one air traffic controller can handle ranges between 10 and
20, depending of the complexity of the traffic pattern. Over a
decade ago, it had been suggested by \cite{RTCA1995} that this
limitation of the air traffic controller can be solved by moving
the responsibility of conflict resolution from the air traffic
controller to the pilots. Since then this airborne self separation
idea has received a lot of research attention. Nevertheless, it
still is unknown how much more air traffic can safely be
accommodated under a well designed airborne self separation way of
working. In order to add to the solution of this debate, a series
of large European studies towards solving this question have been
started under the name HYBRIDGE \cite{HYBRIDGEProject} and iFly
\cite{IFLYProject} respectively. The way of working is to first
develop a well defined SDCPN model of the airborne self separation
concept of operation to be analysed, e.g.\
\cite{EverdijKlompstraBlomKleinObbink2006}. Subsequently this
SDCPN model is further analysed using a bisimilar HSDE and hybrid
automation formal model representation
\cite{BlomKrystulBakkerKlompstraKleinObbink2007,BlomBakkerKrystul2009a},
in which powerful stochastic analysis theory is exploited for the
speeding up of Monte Carlo simulations. Using this approach,
\cite{BlomKleinObbinkBakker2009} have shown that the first
generation of airborne self separation concept designs falls short
in safely accommodating higher air traffic demand than
conventional ATM can. The feedback of this finding to advanced air
traffic concept designers triggered the development of more
advanced airborne self separation concept of operation, e.g.\ see
\cite{IFLYProject}.


\begin{thebibliography}{10}
\providecommand{\bibitemstart}[1]{\bibitem{#1}}
\providecommand{\bibitemend}{}
\providecommand{\bibliographystart}{}
\providecommand{\bibliographyend}{}
\providecommand{\url}[1]{\texttt{#1}}
\providecommand{\urlprefix}{Available at }
\providecommand{\bibinfo}[2]{#2}
\bibliographystart

\bibitemstart{Blom2003}
\bibinfo{author}{H.A.P. Blom} (\bibinfo{year}{2003}):
  \emph{\bibinfo{title}{Stochastic hybrid processes with hybrid jumps}}.
\newblock In: {\sl \bibinfo{booktitle}{Proceedings {IFAC} conference on
  analysis and design of hybrid system ({ADHS}), Saint-Malo, Brittany,
  France}}. pp. \bibinfo{pages}{361--365}.
\bibitemend

\bibitemstart{BlomBakkerEverdijPark2003b}
\bibinfo{author}{H.A.P. Blom}, \bibinfo{author}{G.J. Bakker},
  \bibinfo{author}{M.H.C. Everdij} \& \bibinfo{author}{M.N.J. {Van der Park}}
  (\bibinfo{year}{2003}): \emph{\bibinfo{title}{Stochastic analysis background
  of accident risk assessment for air traffic management}}.
\newblock \bibinfo{type}{HYBRIDGE Report}, \bibinfo{institution}{D2.2}.
\newblock \bibinfo{note}{{h}ttp://hosted.nlr.nl/public/hosted-sites/hybridge/}.
\bibitemend

\bibitemstart{BlomBakkerKrystul2009a}
\bibinfo{author}{H.A.P. Blom}, \bibinfo{author}{G.J. Bakker} \&
  \bibinfo{author}{J.~Krystul} (\bibinfo{year}{2009}):
  \emph{\bibinfo{title}{Rare event estimation for a large scale stochastic
  hybrid system with air traffic application}}.
\newblock In: \bibinfo{editor}{G.~Rubino} \& \bibinfo{editor}{B.~Tuffin},
  editors: {\sl \bibinfo{booktitle}{Rare event simulation using Monte Carlo
  methods}},  \bibinfo{volume}{forthcoming}. \bibinfo{publisher}{J.Wiley}.
\bibitemend

\bibitemstart{BlomKleinObbinkBakker2009}
\bibinfo{author}{H.A.P. Blom}, \bibinfo{author}{B.~{Klein Obbink}} \&
  \bibinfo{author}{G.J. Bakker} (\bibinfo{year}{2009}):
  \emph{\bibinfo{title}{Simulated collision risk of an uncoordinated airborne
  self separation concept of operation}}.
\newblock {\sl \bibinfo{journal}{{ATC} Quarterly}} \bibinfo{volume}{17}, pp.
  \bibinfo{pages}{63--93}.
\bibitemend

\bibitemstart{BlomKrystulBakkerKlompstraKleinObbink2007}
\bibinfo{author}{H.A.P. Blom}, \bibinfo{author}{J.~Krystul},
  \bibinfo{author}{G.J. Bakker}, \bibinfo{author}{M.B. Klompstra} \&
  \bibinfo{author}{B.~{Klein Obbink}} (\bibinfo{year}{2007}):
  \emph{\bibinfo{title}{Free flight collision risk estimation by sequential
  {M}onte {C}arlo simulation}}.
\newblock In: \bibinfo{editor}{C.G. Cassandras} \&
  \bibinfo{editor}{J.~Lygeros}, editors: {\sl \bibinfo{booktitle}{Stochastic
  hybrid systems: recent developments and research trends}}, {\sl
  \bibinfo{series}{Control engineering}}~\bibinfo{volume}{24},
  chapter~\bibinfo{chapter}{10}. \bibinfo{publisher}{Taylor \& Francis Group /
  {CRC} Press}, pp. \bibinfo{pages}{247--281}.
\bibitemend

\bibitemstart{BujorianuLygeros2004cdc}
\bibinfo{author}{M.L. Bujorianu} \& \bibinfo{author}{J.~Lygeros}
  (\bibinfo{year}{2004}): \emph{\bibinfo{title}{General stochastic hybrid
  systems: modelling and optimal control}}.
\newblock In: {\sl \bibinfo{booktitle}{Proceedings 43rd conference on decision
  and control ({CDC}), Nassau, Bahamas}}.
\bibitemend

\bibitemstart{BujorianuLygeros2006}
\bibinfo{author}{M.L. Bujorianu} \& \bibinfo{author}{J.~Lygeros}
  (\bibinfo{year}{2006}): \emph{\bibinfo{title}{Toward a general theory of
  stochastic hybrid systems}}.
\newblock In: \bibinfo{editor}{H.A.P. Blom} \& \bibinfo{editor}{J.~Lygeros},
  editors: {\sl \bibinfo{booktitle}{Stochastic hybrid systems: theory and
  safety critical applications}}, {\sl \bibinfo{series}{Lectures notes in
  control and information sciences {(LNCIS)}}} \bibinfo{volume}{337}.
  \bibinfo{publisher}{Springer}, pp. \bibinfo{pages}{3--30}.
\bibitemend

\bibitemstart{BujorianuLygerosBujorianu2005}
\bibinfo{author}{M.L. Bujorianu}, \bibinfo{author}{J.~Lygeros} \&
  \bibinfo{author}{M.C. Bujorianu} (\bibinfo{year}{2005}):
  \emph{\bibinfo{title}{Different approaches on bisimulation for stochastic
  hybrid systems}}.
\newblock In: \bibinfo{editor}{M.~Morari} \& \bibinfo{editor}{L.~Thiele},
  editors: {\sl \bibinfo{booktitle}{Proceedings 8th international workshop on
  hybrid systems: computation and control ({HSCC}), Z{\"u}rich, Switzerland}},
  {\sl \bibinfo{series}{Lecture notes in computer science ({LNCS})}}
  \bibinfo{volume}{3414}. pp. \bibinfo{pages}{198--214}.
\bibitemend

\bibitemstart{Davis1993}
\bibinfo{author}{M.H.A. Davis} (\bibinfo{year}{1993}):
  \emph{\bibinfo{title}{Markov models and optimization}}, {\sl
  \bibinfo{series}{Monographs on statistics and applied
  probability}}~\bibinfo{volume}{49}.
\newblock \bibinfo{publisher}{Chapman and Hall}, \bibinfo{address}{London}.
\bibitemend

\bibitemstart{Elliott1982}
\bibinfo{author}{R.J. Elliott} (\bibinfo{year}{1982}):
  \emph{\bibinfo{title}{Stochastic calculus and applications}}, {\sl
  \bibinfo{series}{Applications of mathematics: Stochastic modelling and
  applied probability}}~\bibinfo{volume}{18}.
\newblock \bibinfo{publisher}{Springer-Verlag}.
\bibitemend

\bibitemstart{ElliottAggounMoore1995}
\bibinfo{author}{R.J. Elliott}, \bibinfo{author}{L.~Aggoun} \&
  \bibinfo{author}{J.B. Moore} (\bibinfo{year}{1995}):
  \emph{\bibinfo{title}{Hidden {M}arkov models: estimation and control}}, {\sl
  \bibinfo{series}{Applications of mathematics: stochastic modelling and
  applied probability}}~\bibinfo{volume}{29}.
\newblock \bibinfo{publisher}{Springer-Verlag}.
\bibitemend

\bibitemstart{EthierKurtz1986}
\bibinfo{author}{S.N. Ethier} \& \bibinfo{author}{T.G. Kurtz}
  (\bibinfo{year}{1986}): \emph{\bibinfo{title}{Markov processes,
  characterization and convergence}}.
\newblock Wiley series in probability and mathematical statistics.
  \bibinfo{publisher}{John Wiley \& Sons}, \bibinfo{address}{New York}.
\bibitemend

\bibitemstart{EverdijBlom2003}
\bibinfo{author}{M.H.C. Everdij} \& \bibinfo{author}{H.A.P. Blom}
  (\bibinfo{year}{2003}): \emph{\bibinfo{title}{{P}etri nets and hybrid state
  {M}arkov processes in a power-hierarchy of dependability models}}.
\newblock In: {\sl \bibinfo{booktitle}{Proceedings {IFAC} conference on
  analysis and design of hybrid system ({ADHS}), Saint-Malo, Brittany,
  France}}. pp. \bibinfo{pages}{355--360}.
\bibitemend

\bibitemstart{EverdijBlom2005}
\bibinfo{author}{M.H.C. Everdij} \& \bibinfo{author}{H.A.P. Blom}
  (\bibinfo{year}{2005}): \emph{\bibinfo{title}{Piecewise deterministic
  {M}arkov processes represented by dynamically coloured {P}etri nets}}.
\newblock In: \bibinfo{editor}{S.~Jacka}, editor: {\sl
  \bibinfo{booktitle}{Stochastics: an international journal of probability and
  stochastic processes}},  \bibinfo{volume}{77, number 1}.
  \bibinfo{publisher}{Taylor \& Francis}, pp. \bibinfo{pages}{1--29}.
\bibitemend

\bibitemstart{EverdijBlom2006}
\bibinfo{author}{M.H.C. Everdij} \& \bibinfo{author}{H.A.P. Blom}
  (\bibinfo{year}{2006}): \emph{\bibinfo{title}{Hybrid {P}etri nets with
  diffusion that have into-mappings with generalised stochastic hybrid
  processes}}.
\newblock In: \bibinfo{editor}{H.A.P. Blom} \& \bibinfo{editor}{J.~Lygeros},
  editors: {\sl \bibinfo{booktitle}{Stochastic hybrid systems: theory and
  safety critical applications}}, {\sl \bibinfo{series}{Lectures notes in
  control and information sciences {(LNCIS)}}} \bibinfo{volume}{337}.
  \bibinfo{publisher}{Springer}, pp. \bibinfo{pages}{31--63}.
\bibitemend

\bibitemstart{EverdijBlom2009}
\bibinfo{author}{M.H.C. Everdij} \& \bibinfo{author}{H.A.P. Blom}
  (\bibinfo{year}{2010}): \emph{\bibinfo{title}{Hybrid state Petri nets which
  have the analysis power of stochastic hybrid systems and the formal
  verification power of automata}}.
\newblock In: \bibinfo{editor}{P.~Pawlewski}, editor: {\sl
  \bibinfo{booktitle}{Petri nets: Applications}},
  chapter~\bibinfo{chapter}{12}. \bibinfo{publisher}{InTech}, pp.
  \bibinfo{pages}{227--252}.
\bibitemend

\bibitemstart{EverdijKlompstraBlomKleinObbink2006}
\bibinfo{author}{M.H.C. Everdij}, \bibinfo{author}{M.B. Klompstra},
  \bibinfo{author}{H.A.P. Blom} \& \bibinfo{author}{B.~{Klein Obbink}}
  (\bibinfo{year}{2006}): \emph{\bibinfo{title}{Compositional specification of
  a multi-agent system by stochastically and dynamically coloured {P}etri
  nets}}.
\newblock In: \bibinfo{editor}{H.A.P. Blom} \& \bibinfo{editor}{J.~Lygeros},
  editors: {\sl \bibinfo{booktitle}{Stochastic hybrid systems: theory and
  safety critical applications}}, {\sl \bibinfo{series}{Lectures notes in
  control and information sciences {(LNCIS)}}} \bibinfo{volume}{337}.
  \bibinfo{publisher}{Springer}, pp. \bibinfo{pages}{325--350}.
\bibitemend

\bibitemstart{HYBRIDGEProject}
\bibinfo{author}{{HYBRIDGE}} (\bibinfo{year}{2002}).
\newblock \emph{\bibinfo{title}{{EC} project description}}.
\newblock
  \bibinfo{note}{{h}ttp://hosted.nlr.nl/public/hosted-sites/\-hybridge}.
\bibitemend

\bibitemstart{IFLYProject}
\bibinfo{author}{{iFly}} (\bibinfo{year}{2007}).
\newblock \emph{\bibinfo{title}{{EC} project description}}.
\newblock \bibinfo{note}{{h}ttp://iFly.nlr.nl}.
\bibitemend

\bibitemstart{Krystul2006}
\bibinfo{author}{J.~Krystul} (\bibinfo{year}{2006}):
  \emph{\bibinfo{title}{Modelling of stochastic hybrid systems with
  applications to accident risk assessment}}.
\newblock \bibinfo{type}{Ph.D. thesis}, \bibinfo{school}{University of Twente},
  \bibinfo{address}{The Netherlands}.
\bibitemend

\bibitemstart{KrystulBlomBagchi2007}
\bibinfo{author}{J.~Krystul}, \bibinfo{author}{H.A.P. Blom} \&
  \bibinfo{author}{A.~Bagchi} (\bibinfo{year}{2007}):
  \emph{\bibinfo{title}{Stochastic hybrid systems}}, chapter
  \bibinfo{chapter}{2: Stochastic differential equations on hybrid state
  spaces}, pp. \bibinfo{pages}{15--45}.
\newblock Number~\bibinfo{number}{24} in \bibinfo{series}{Control engineering
  series}. \bibinfo{publisher}{Taylor and Francis / CRC Press}.
\bibitemend

\bibitemstart{RTCA1995}
\bibinfo{author}{{RTCA}} (\bibinfo{year}{1995}): \emph{\bibinfo{title}{Final
  report of {RTCA} {T}ask {F}orce 3; Free Flight implementation}}.
\newblock \bibinfo{type}{Final}, \bibinfo{institution}{RTCA Inc.},
  \bibinfo{address}{Washington DC}.
\bibitemend

\bibitemstart{VanderSchaft2004}
\bibinfo{author}{A.J. {Van der Schaft}} (\bibinfo{year}{2004}):
  \emph{\bibinfo{title}{Equivalence of dynamical systems by bisimulation}}.
\newblock {\sl \bibinfo{journal}{{IEEE} transactions on automatic control}}
  \bibinfo{volume}{49}(\bibinfo{number}{12}), pp. \bibinfo{pages}{2160--2172}.
\bibitemend

\bibliographyend
\end{thebibliography}

\end{document}